\documentclass{article}

\usepackage[dblblindworkshop,final,nonatbib]{neurips_2025}
\workshoptitle{Machine Learning and the Physical Sciences}

\usepackage[utf8]{inputenc} % allow utf-8 input
\usepackage[T1]{fontenc}    % use 8-bit T1 fonts
\usepackage{hyperref}       % hyperlinks
\usepackage{url}            % simple URL typesetting
\usepackage{booktabs}       % professional-quality tables
\usepackage{amsfonts}       % blackboard math symbols
\usepackage{nicefrac}       % compact symbols for 1/2, etc.
\usepackage{microtype}      % microtypography
\usepackage{xcolor}         % colors
\usepackage{graphicx}
\usepackage{amsmath}
\usepackage{multirow}
\usepackage{subfig}

\usepackage{amssymb}% http://ctan.org/pkg/amssymb
\usepackage{pifont}% http://ctan.org/pkg/pifont
\newcommand{\cmark}{\ding{51}}%
\newcommand{\xmark}{\ding{55}}%

\title{
The Pareto frontier of resilient jet tagging
}

\author{%
  Rikab~Gambhir\thanks{gambhirb@ucmail.uc.edu} \\
  University of Cincinnati / IAIFI
  \And
  Matt~LeBlanc\thanks{matt\_leblanc@brown.edu} \\
  Brown University / IAIFI
  \And
  Yuanchen~Zhou\thanks{yuanchen\_zhou@brown.edu} \\
  Brown University
}

\begin{document}

\maketitle

%%%%%%%%%%%%%%%%%%%%%%%%%%%%%%%%%%%%%%%%%%%%%%%%%%%%%%%%%%%%%%%%%%%%%%%%%%%%%

\begin{abstract}
  % The abstract paragraph should be indented \nicefrac{1}{2}~inch (3~picas) on
  % both the left- and right-hand margins. Use 10~point type, with a vertical
  % spacing (leading) of 11~points.  The word \textbf{Abstract} must be centered,
  % bold, and in point size 12. Two line spaces precede the abstract. The abstract
  % must be limited to one paragraph.
Classifying hadronic jets using their constituents' kinematic information is a critical task in modern high-energy collider physics.
Often, classifiers are designed by targeting the best performance using metrics such as accuracy, AUC, or rejection rates.
However, the use of a single metric can lead to the use of architectures that are more model-dependent than competitive alternatives, leading to potential uncertainty and bias in analysis.
We explore such trade-offs and demonstrate the consequences of using networks with high performance metrics but low resilience.
%
% This paper explores such trade-offs, and explores knowledge distillation as a way to improve the performance of classifiers while minimizing their model-dependence.
\end{abstract}

%%%%%%%%%%%%%%%%%%%%%%%%%%%%%%%%%%%%%%%%%%%%%%%%%%%%%%%%%%%%%%%%%%%%%%%%%%%%%
\section{Introduction}

When strongly-interacting quarks and gluons are produced by high-energy particle collisions at colliders like the Large Hadron Collider (LHC), they shower and hadronize, creating a collimated `jet' of particles in the final state that is imprinted with some properties of the originating particle~\cite{Salam:2010nqg}.
Classification, or \emph{tagging}, of these jets based on their \emph{substructure} has become a critical task at the Large Hadron Collider (LHC), where many studies require doing so to extract maximal information from the data~\cite{Marzani:2019hun}.
Jet tagging has become the quintessential proving grounds for Artificial Intelligence / Machine Learning (AI/ML) algorithms at the LHC: state-of-the-art transformer and graph-based architectures~\cite{ATL-PHYS-PUB-2023-032,ATL-PHYS-PUB-2023-020,ATL-PHYS-PUB-2023-017,ATL-PHYS-PUB-2022-039,CMS-JME-18-002} are significantly more performant than earlier approaches~\cite{CMS-JME-13-006,PERF-2015-03,PERF-2015-04,JETM-2018-03,JETM-2020-02}.

Economists say, ``When a measure becomes a target, it ceases to be a good measure.''~\cite{10.2307/2232324}
While the accuracy of an ML/AI classifier, often measured by `AUC' (area under the ROC curve), is a critical benchmark, fixation on this quantity can lead to sub-optimal outcomes in analyses.
As model complexity increases, they can become susceptible to learning idiosyncrasies of the simulated training sample rather than genuine generalizable physics information.
This has been studied by ATLAS~\cite{JETM-2023-06}, which showed that classifiers are more susceptible to uncertainties related to physics modeling than those related to detector effects.
Similar studies have explored solutions to generalizability~\cite{Butter:2022xyj}.

In this work, we evaluate architectures that are often used for tagging in terms of their AUC and their simulation model-dependence, or`resilience'~\cite{Soyez:2018opl,Proceedings:2018jsb}.
Models with varying complexities were trained and tested on \emph{different} Monte Carlo (MC) simulated datasets; then used to construct the `Pareto frontier'~\cite{10.1007/978-3-540-88908-3_14} of AUC \emph{vs.} resilience.
We studied knowledge distillation from complex to simple network types as one approach to beating the Pareto frontier, as it had previously been reported to improve the tagging performance of simple networks~\cite{Liu:2023dio}.
While we were unsuccessful in overcoming the frontier set without Distillation techniques, studying resilience in the context of distillation was nevertheless of interest.
Finally, we perform a case study to demonstrate the risk of biasing downstream parameter estimation tasks when using models with low resilience.
We advocate for a holistic approach to classifier design that includes multiple benchmarks, suited to the application.

%%%%%%%%%%%%%%%%%%%%%%%%%%%%%%%%%%%%%%%%%%%%%%%%%%%%%%%%%%%%%%%%%%%%%%%%%%%%%
\section{Methodology}\label{sec:methods}

\subsection{Monte Carlo simulated event samples}
 
Two jet classification tasks were considered in these studies: the discrimination of jets initiated by a quark or gluon (`q/g tagging'), and the identification of jets resulting from the hadronic decay of a Lorentz-boosted top quark (`top tagging').
For each of these tasks, a set of MC simulated events generated with \textsc{Pythia}~8~\cite{Sjostrand:2007gs} was used to train the classifiers in a fully-supervised manner.
\textsc{Pythia} samples used the default Monash set of tuned parameters~\cite{Skands:2014pea} in all cases.
Alternative samples of the same processes were also generated with \textsc{Herwig}~7~\cite{Bellm:2017bvx}\footnote{For q/g (top), samples are generated with \textsc{Pythia} version 8.226 (8.331) and \textsc{Herwig} version 7.1.4 (7.3.0).}, to enable quantification of the resilience as the AUC \%-difference between testing on the nominal and alternative sample.\footnote{There are many potential ways to define resilience.
Just as AUC does not capture all important aspects of classification, so too does this particular definition of resilience.}
All jets, regardless of their size, are reconstructed with \textsc{FastJet}~\cite{fastjet} and filtered to have a transverse momentum ($p_{\mathrm{T}}$) between 500-550~GeV.
No detector simulation is applied.

For q/g tagging studies, the simulated event samples from Refs.~\cite{Komiske:2018cqr} were used, which are freely available on the CERN Zenodo platform~\cite{komiske_2019_3164691,pathak_2019_3066475}.
These samples consist of one million anti-$k_t$ $R=0.4$ jets from each of the $Z+q$ or $Z+g$ processes, where the $Z$ boson decays into neutrinos.
The boosted top tagging studies were performed using a new set of samples, which consist of one million hadronically-decaying signal top quark jets were selected from top pair-production processes in each generator, reconstructed as ungroomed anti-$k_t$ jets with radius $R=0.8$.
One million mixed $R=0.8$ $q/g$ background jets for this task with the same kinematic selection were produced from a dijet process, with a physical q/g mixture fraction.
These samples have been made publicly available on the CERN Zenodo platform~\cite{rikab_top_qcd}, and provide a convenient, interoperable extension of the existing $q/g$ jets samples due to their shared format.

\subsection{Model architectures surveyed}\label{sec:models}

We have surveyed a representative selection of architectures that are either currently used or have recently been used in physics analysis at the LHC, and summarize the setups studied in Table~\ref{tab:tagger-hyperparams}.
All networks were trained in a fully-supervised manner for 500 epochs,\footnote{This number was chosen arbitrarily, the early stopping condition causes training to terminate in between 30-100 epochs in most cases.}, using early stopping and a patience of 10 epochs.
The default \textsc{Adam}~\cite{kingma2017adammethodstochasticoptimization} optimizer was used, with a learning rate of $0.001$.
The MC samples were split such that 75\% of the events were used for training, and 12.5\% each were used for testing \& validation.
Each network was given only particle-level kinematic information (constituent $p_T$, pseudorapidity $\eta$, and azimuthal angle $\phi$) as input.

% \begin{itemize}
% \item \textbf{Expert features}: Handcrafted physics observables often used in tagging, \emph{i.e.} particle multiplicities or jet angularities~\cite{Larkoski:2014pca}. Different values of $\beta$ were studied for angularities, and multiplicities with various requirements on $p_{\text{T}}$ and particle charges were used.
% \item \textbf{Deep Neural Networks (DNNs)}: a fully-connected multi-layer perceptron. Networks with 2-10 hidden layers and 1-300 neurons per-hidden-layer were studied.
% \item \textbf{Particle-Flow Networks (PFNs)}: a permutation-invariant deep-sets-based architecture~\cite{Komiske:2018cqr}. Latent dimensions $\ell=1-1024$ and $\phi$ sizes of between $50-500$ nodes were studied.
% \item \textbf{Energy-Flow Networks (EFNs)}: an IRC-safe variant of the PFN~\cite{Komiske:2018cqr}. The same hyperparameters were scanned as for PFNs.
% % \item \textbf{Particle Net (PNet)}: a graph-based architecture~\cite{Qu:2019gqs}.
% \item \textbf{Particle Transformer (ParT)}: a transformer-based architecture~\cite{pmlr-v162-qu22b}. Setups with 2, 4 and 8 attention heads were used. 
% \end{itemize}

\begin{table}[h]
  \caption{
    An overview of the jet tagging architectures surveyed in this study.
    For each tagger, the key hyperparameters that were scanned to explore the performance-resilience trade-off are listed.
  }
  \label{tab:tagger-hyperparams}
  \centering
  \begin{tabular}{llc}
    \toprule
    \textbf{Architecture} & \textbf{Hyperparameters Scanned} & \textbf{Reference} \\
    \midrule
    Expert Features & Angularities: $\beta$ values & \cite{Larkoski:2014pca} \\
    & Multiplicities: $p_{\text{T}}$ cuts, charge req. & \\
    \midrule
    Deep Neural Networks (DNNs) & Hidden layers: 2--10 & \cite{Komiske:2018cqr} \\
    & Neurons per layer: 1--300 & \\
    \midrule
    Particle-Flow Networks (PFNs) & Latent dimension $\ell$: 1--1024 & \cite{Komiske:2018cqr} \\
    & $\Phi$ network nodes: 50--500 & \\
    \midrule
    Energy-Flow Networks (EFNs) & Same as PFNs & \cite{Komiske:2018cqr} \\
    \midrule
    Particle Transformer (ParT) & Attention heads: 2, 4, 8 & \cite{pmlr-v162-qu22b} \\
    \bottomrule
  \end{tabular}
\end{table}

\section{Results}\label{sec:results}

\subsection{Pareto Frontier}

\begin{figure}[htb]
    \centering
    \subfloat[\label{fig:pareto:a}]{ 
    \includegraphics[width=0.48\linewidth]{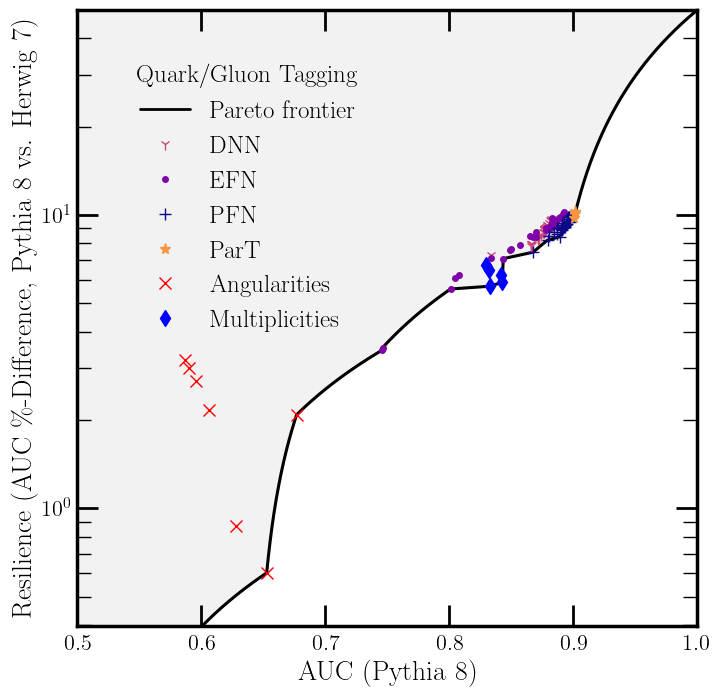}
    }
    \subfloat[\label{fig:pareto:b}]{
    \includegraphics[width=0.49\linewidth]{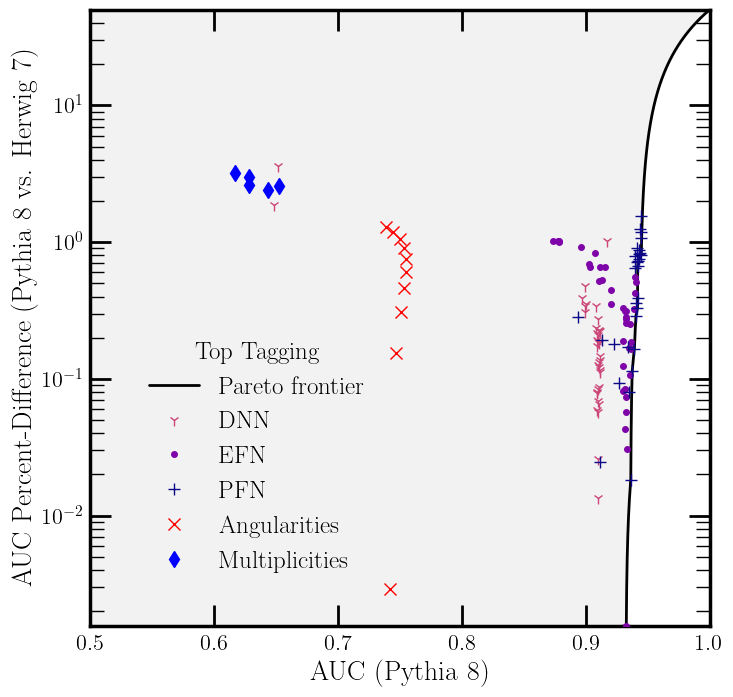}
    }
    \caption{The Pareto frontier for (a) q/g tagging, (b) top tagging tasks.
    The AUC of models trained with \textsc{Pythia} samples is plotted \emph{vs.} the resilience, defined as the percent difference in AUC evaluated on the \textsc{Pythia} and \textsc{Herwig} test samples. 
    The various markers denote different classifiers: DNN (triangles), EFN (circles), PFN (pluses), ParT (stars), Angularities (crosses) and Multiplicities (diamonds).
    The shaded grey region is Pareto-excluded.}
    \label{fig:pareto_frontier}
\end{figure}

Figure~\ref{fig:pareto_frontier} shows the classifier AUC \emph{vs.} resilience (AUC \%-difference for Pythia vs. Herwig samples), for each of the models trained and for each set of hyperparameters considered: `optimal' performance is at the lower-right corner of the figure.
The Pareto frontier connecting models that optimize the AUC-resiliency tradeoff is highlighted, and models in the shaded region are Pareto-excluded.

The Pareto frontier shows that more ``complex'' models (e.g. ParT) do achieve a higher raw performance, but at the cost of resiliency.
On the other hand, simpler models based on physical principles, such as EFNs or mere expert features like angularities, are more robust.
In particular, multiplicities are powerful $q/g$ discriminants that appear to move the Pareto frontier forward of that which would be drawn from EFNs alone, despite their lacking IRC-safety.
For top tagging, vertical columns of different observable/network types strongly discourage the use of unnecessarily complex networks.

\subsection{Knowledge Distillation}

% Given a boundary, we should endeavor to overcome it.
%
In an attempt to overcome the Pareto frontier and make models better along both AUC and resiliency, we tried Knowledge Distillation: a complex `teacher' model is used to train a less complex `student' model~\cite{hinton2015distillingknowledgeneuralnetwork}.
Ultimately, this approach was unsuccessful in overcoming the Pareto frontier, but interesting observations made during the study motivate us to document it here.
The PFN with $\ell=128$ and 250 dense nodes per hidden layer was used as the teacher for this study, while various DNN and EFN models were used as students.
The training procedure in Section~\ref{sec:models} was modified such that the students were trained instead using the teacher's prediction as `soft labels' by minimizing the KL-divergence between the predictions of the teacher and student models per-batch.
Forms of L1 and L2 regularization were tested~\cite{51791361-8fe2-38d5-959f-ae8d048b490d,a92f3c16-7c6e-31d3-b403-82d2b0a469e4,tikhonov1977solutions}, which were expected to help prevent students over-fitting the teacher's response.
No significant change in the outcome was found, and so for simplicity the final version of the study therefore did not apply a regularization method.

\begin{figure}[htb]
    \centering
     \subfloat[\label{fig:distill:a}]{
        \includegraphics[width=0.49\linewidth]{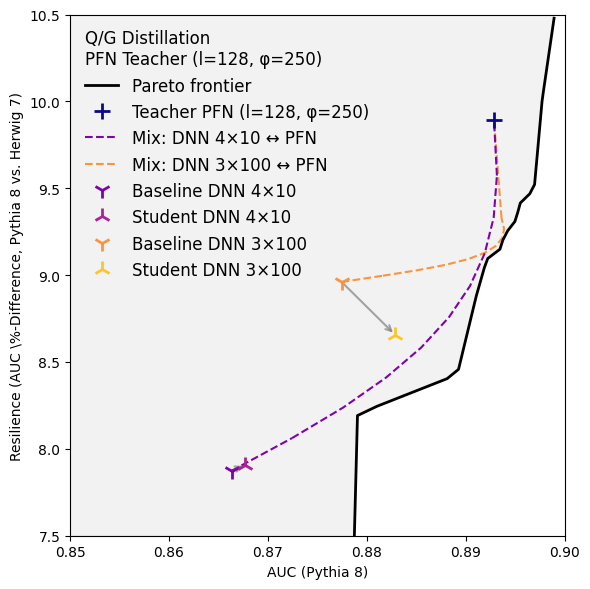}
     }
     \subfloat[\label{fig:distill:b}]{
        \includegraphics[width=0.49\linewidth]{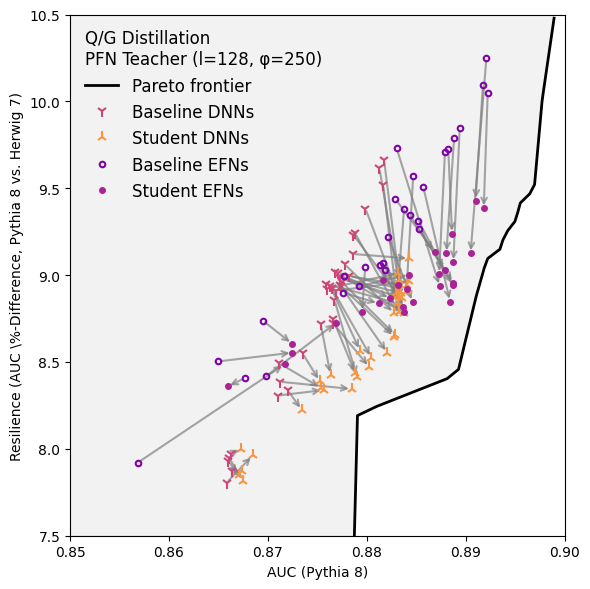}
     }
    \caption{(a) Results of training two student DNNs via distillation from a teacher PFN. (b) Summary of distillation training from a teacher PFN to all DNNs and EFNs in the study.}
    \label{fig:distillation:a}
\end{figure}

The results for a representative pair of student models are shown in the AUC-Resiliance plane in Figure~\ref{fig:distill:a}, along with the teacher model and `baseline' models whose architectures match the students, but that are not trained using distillation.
The contour between the baseline and teacher models is also drawn: it is obtained by performing inference with a linear combination of the two models on the test set that varies from pure-teacher to pure-baseline, in 10\% steps.
The student models beat this contour, demonstrating non-trivial improvement: the AUC of the model increases more than its resilience degrades.
However, when models that are closer to the Pareto frontier are selected for distillation, the observed improvement is reduced.
The results of distilling into the many DNN and EFN models we study is shown in Figure~\ref{fig:distill:b}: while many students improve, none surpass the existing frontier.

\vspace{-5pt}
\vspace{-5pt}

\subsection{Case Study: determining q/g fractions}\label{sec:qgfrac}

For a realistic downstream analysis task, a less accurate but more resilient classifier can ultimately yield a less biased physics result.
We illustrate this with a case study, where the flavor \emph{mixture fraction} $\kappa$ of a mixed sample of quark and gluon jets is estimated using two PFNs located on the Pareto frontier: a small, resilient PFN ($\ell=8$, 50 nodes per hidden layer) and a large network with a higher AUC ($\ell=128$, 250 nodes per hidden layer).
Given a classifier, one can extract the per-event q/g likelihood ratio $\ell(x)$ via the Neyman Pearson Lemma~\cite{neyman-pearson}, from which the $\kappa$ maximum likelihood inferences may be extracted.
The flavor composition of two mixed jet samples with respective quark-initiated jet fractions of $50\%$ and $25\%$ is estimated with both networks, for jets from either the \textsc{Pythia} or \textsc{Herwig} samples; the results are tabulated in Table~\ref{tab:kappa-estimation}.
The experiment is re-run 10 times with re-trained networks as a proxy for confidence interval estimation.

Both the large and small networks are able to accurately recover the mixture fraction when samples are constructed with \textsc{Pythia} jets.
However, when classifiers trained with \textsc{Pythia} jets are used to estimate the mixture fraction in a sample of \textsc{Herwig} jets (used as pseudodata sample in this study), the inferred $\kappa$ value is biased.
The extraction process can be calibrated by reweighting using a \emph{second} set of classifiers that distinguish between \textsc{Pythia} and \textsc{Herwig}.
This classifier approximates the likelihood ratio between the two classes, allowing one to reweight one class of samples to be statistically identical from the other class (modulo reweighting uncertainties)\footnote{%
In principle, one can extract reweighting uncertainties and obtain confidence intervals using a method such as WiFi ensembles~\cite{Benevedes:2025nzr}, but for simplicity we do not do this here.
%
% These weights would then used as the $w_i$ in \eqref{eq:kappa}.
}.
The PFN models used for the reweighting are identical in architecture to those used for classification.
Following calibration, we see from Table~\ref{tab:kappa-estimation} that the less resilient model is still biased: the inferred $\hat\kappa$ values are not statistically consistent with the true $\kappa$ values.
The more resilient model is unbiased (within $2\sigma$) following the calibration procedure, despite its naively worse performance according to the AUC. 
The conclusions of this study may apply broadly: any perturbation of the correlation structure of a sample with respect to the training set (\emph{e.g.} fast \emph{vs.} full detector simulation) may result in such a bias for parameter extractions depending on a unresilient classifier.
This is particularly relevant to substructure, as predictions are known to differ from each other and from measurement~\cite{STDM-2017-04,STDM-2017-33,STDM-2017-16,STDM-2018-57,ATLAS:2024wrd,ATLAS:2025qtv,CMS-SMP-20-010,CMS:2023lpp}.

\begin{table}[htb]
  \caption{
    Summary of the q/g mixture fraction ($\kappa$) estimation case study.
    Reported uncertainties are determined from the standard error from multiple pseudo-experiments.
  }
  \label{tab:kappa-estimation}
  \centering
  \begin{tabular}{lccccc}
    \toprule
    & & \textit{Pythia 8} & \multicolumn{2}{c}{\textit{Herwig 7}} \\
    \cmidrule(lr){3-3} \cmidrule(lr){4-5}
    \textbf{Classifier} & \textbf{True $\kappa$} & \textbf{Inferred $\mathbb{E}[\hat\kappa]$} & \textbf{Inferred $\mathbb{E}[\hat\kappa]$} & \textbf{Calibrated $\mathbb{E}[\hat\kappa]$} & \textbf{Result} \\
    \midrule
    Large PFN         & 0.50 & $0.490 \pm 0.005$ & $0.296 \pm 0.007$ & $0.529 \pm 0.006$ & Biased \xmark \\
      & 0.25 & $0.253 \pm 0.005$ & $0.125 \pm 0.005$ & $0.305 \pm 0.006$ & Biased \xmark \\
    \midrule
    Small PFN         & 0.50 & $0.504 \pm 0.013$ & $0.336 \pm 0.016$ & $0.478 \pm 0.017$ & Unbiased \cmark \\
     & 0.25 & $0.258 \pm 0.013$ & $0.157 \pm 0.014$ & $0.268 \pm 0.013$ & Unbiased \cmark \\
    \bottomrule
  \end{tabular}
\end{table}

%%%%%%%%%%%%%%%%%%%%%%%%%%%%%%%%%%%%%%%%%%%%%%%%%%%%%%%%%%%%%%%%%%%%%%%%%%%%%
\section{Concluding remarks}\label{sec:conclusion}

There is a clear trade-off between classifier performance and resilience, which we have visualized in these studies as a Pareto frontier.
We have found that the complexity of a given model is a primary driver along the frontier, and that suboptimal model architectures can be improved with more sophisticated approaches to training such as knowledge distillation.
Ultimately, the choice of a classifier that is not resilient can lead to suboptimal performance and increased bias in downstream tasks, even if the model is more accurate than others, motivating a more holistic approach to classifier development that includes multiple benchmarks.
For example, the inference time of classifiers deployed in online data-taking should also be strongly considered during the design stage.

% \emph{The utilitarian cares only for maximizing AUC, and so overtrains a look-up table.}

% \emph{The deontologist cares only for resilient theory, and so classifies with angularities at leading-log.}

%%%%%%%%%%%%%%%%%%%%%%%%%%%%%%%%%%%%%%%%%%%%%%%%%%%%%%%%%%%%%%%%%%%%%%%%%%%%%

\begin{ack}

MLB and YZ gratefully acknowledge the support of the Brown University Department of Physics, and would like to thank the MIT Center for Theoretical Physics and IAIFI for their frequent hospitality.
They are supported by the U.S. Department of Energy, Office of Science, Office of High Energy Physics under Award Number DE-SC0026285.
RG is supported by the National Science Foundation under Cooperative Agreement PHY-2019786 (The NSF AI Institute for Artificial Intelligence and Fundamental Interactions, \url{http://iaifi.org/}), the NSF grant numbers OAC-2103889, OAC-2411215, and OAC-2417682; and by the U.S. DOE Office of High Energy Physics under grant numbers DE-SC0012567 and DE-SC1019775. 
Finally, the authors would like to thank Sean Benevedes and Jennifer Roloff for their prompt feedback on the draft, and to express their gratitude to Huilin Qu for his prompt assistance when setting up ParT, and to Tilman Plehn for his useful feedback at the 2024 IAIFI Workshop.

\end{ack}

%%%%%%%%%%%%%%%%%%%%%%%%%%%%%%%%%%%%%%%%%%%%%%%%%%%%%%%%%%%%%%%%%%%%%%%%%%%%%

{
\small

\bibliographystyle{ieeetr}
\bibliography{refs.bib,PubNotes.bib,ATLAS.bib,CMS.bib}

@Article{PERF-2015-03,
    author         = "{ATLAS Collaboration}",
    title          = "{Identification of boosted, hadronically decaying \(W\) bosons and comparisons with ATLAS data taken at \(\sqrt{s} = 8\,\text{TeV}\)}",
    journal        = "Eur. Phys. J. C",
    volume         = "76",
    year           = "2016",
    pages          = "154",
    doi            = "10.1140/epjc/s10052-016-3978-z",
    reportNumber   = "CERN-PH-EP-2015-204",
    eprint         = "1510.05821",
    archivePrefix  = "arXiv",
    primaryClass   = "hep-ex",
}

@Article{PERF-2015-04,
    author         = "{ATLAS Collaboration}",
    title          = "{Identification of high transverse momentum top quarks in \(pp\) collisions at \(\sqrt{s} = 8\,\text{TeV}\) with the ATLAS detector}",
    journal        = "JHEP",
    volume         = "06",
    year           = "2016",
    pages          = "093",
    doi            = "10.1007/JHEP06(2016)093",
    reportNumber   = "CERN-EP-2016-010",
    eprint         = "1603.03127",
    archivePrefix  = "arXiv",
    primaryClass   = "hep-ex",
}

@Article{STDM-2017-04,
    author         = "{ATLAS Collaboration}",
    title          = "{Measurement of the Soft-Drop Jet Mass in \(pp\) Collisions at \(\sqrt{s} = 13\,\text{TeV}\) with the ATLAS detector}",
    journal        = "Phys. Rev. Lett.",
    volume         = "121",
    year           = "2018",
    pages          = "092001",
    doi            = "10.1103/PhysRevLett.121.092001",
    reportNumber   = "CERN-EP-2017-231",
    eprint         = "1711.08341",
    archivePrefix  = "arXiv",
    primaryClass   = "hep-ex",
}

@Article{STDM-2017-16,
    author         = "{ATLAS Collaboration}",
    title          = "{Properties of jet fragmentation using charged particles measured with the ATLAS detector in \(pp\) collisions at \(\sqrt{s} = 13\,\text{TeV}\)}",
    journal        = "Phys. Rev. D",
    volume         = "100",
    year           = "2019",
    pages          = "052011",
    doi            = "10.1103/PhysRevD.100.052011",
    reportNumber   = "CERN-EP-2019-090",
    eprint         = "1906.09254",
    archivePrefix  = "arXiv",
    primaryClass   = "hep-ex",
}

@Article{STDM-2017-33,
    author         = "{ATLAS Collaboration}",
    title          = "{Measurement of soft-drop jet observables in \(pp\) collisions with the ATLAS detector at \(\sqrt{s} = 13\,\text{TeV}\)}",
    journal        = "Phys. Rev. D",
    volume         = "101",
    year           = "2020",
    pages          = "052007",
    doi            = "10.1103/PhysRevD.101.052007",
    reportNumber   = "CERN-EP-2019-269",
    eprint         = "1912.09837",
    archivePrefix  = "arXiv",
    primaryClass   = "hep-ex",
}

@Article{JETM-2018-03,
    author         = "{ATLAS Collaboration}",
    title          = "{Performance of top-quark and \(W\)-boson tagging with ATLAS in Run~2 of the LHC}",
    journal        = "Eur. Phys. J. C",
    volume         = "79",
    year           = "2019",
    pages          = "375",
    doi            = "10.1140/epjc/s10052-019-6847-8",
    reportNumber   = "CERN-EP-2018-192",
    eprint         = "1808.07858",
    archivePrefix  = "arXiv",
    primaryClass   = "hep-ex",
}

@Article{STDM-2018-57,
    author         = "{ATLAS Collaboration}",
    title          = "{Measurement of the Lund Jet Plane Using Charged Particles in \(13\,\text{TeV}\) Proton--Proton Collisions with the ATLAS Detector}",
    journal        = "Phys. Rev. Lett.",
    volume         = "124",
    year           = "2020",
    pages          = "222002",
    doi            = "10.1103/PhysRevLett.124.222002",
    reportNumber   = "CERN-EP-2020-030",
    eprint         = "2004.03540",
    archivePrefix  = "arXiv",
    primaryClass   = "hep-ex",
}

@Article{JETM-2020-02,
    author         = "{ATLAS Collaboration}",
    title          = "{Performance and calibration of quark/gluon-jet taggers using \(140\,\text{fb}^{-1}\) of \(pp\) collisions at \(\sqrt{s} = 13\,\text{TeV}\) with the ATLAS detector}",
    reportNumber   = "CERN-EP-2023-151",
    eprint         = "2308.00716",
    archivePrefix  = "arXiv",
    primaryClass   = "hep-ex",
    doi = "10.1088/1674-1137/acf701",
    journal = "Chin. Phys. C",
    volume = "48",
    number = "2",
    pages = "023001",
    year = "2024"
}

@Article{CMS-JME-13-006,
    author         = "{CMS Collaboration}",
    title          = "{Identification techniques for highly boosted \(W\) bosons that decay into hadrons}",
    journal        = "JHEP",
    volume         = "12",
    year           = "2014",
    pages          = "017",
    doi            = "10.1007/JHEP12(2014)017",
    reportNumber   = "CERN-PH-EP-2014-241",
    eprint         = "1410.4227",
    archivePrefix  = "arXiv",
    primaryClass   = "hep-ex",
}

@Article{CMS-JME-18-002,
    author         = "{CMS Collaboration}",
    title          = "{Identification of heavy, energetic, hadronically decaying particles using machine-learning techniques}",
    journal        = "JINST",
    volume         = "15",
    year           = "2020",
    pages          = "P06005",
    doi            = "10.1088/1748-0221/15/06/P06005",
    reportNumber   = "CERN-EP-2020-037",
    eprint         = "2004.08262",
    archivePrefix  = "arXiv",
    primaryClass   = "hep-ex",
}

@Article{CMS-SMP-20-010,
    author         = "{CMS Collaboration}",
    title          = "{Study of quark and gluon jet substructure in \(Z\)+jet and dijet events from \(pp\) collisions}",
    journal        = "JHEP",
    volume         = "01",
    year           = "2022",
    pages          = "188",
    doi            = "10.1007/JHEP01(2022)188",
    reportNumber   = "CERN-EP-2021-161",
    eprint         = "2109.03340",
    archivePrefix  = "arXiv",
    primaryClass   = "hep-ex",
}

@Booklet{ATL-PHYS-PUB-2022-039,
    author         = "{ATLAS Collaboration}",
    title          = "{Constituent-Based Top-Quark Tagging with the ATLAS Detector}",
    howpublished   = "{ATL-PHYS-PUB-2022-039}",
    url            = "https://cds.cern.ch/record/2825328",
    year           = "2022",
}

@Booklet{ATL-PHYS-PUB-2023-017,
    author         = "{ATLAS Collaboration}",
    title          = "{Tagging boosted \(W\) bosons applying machine learning to the Lund Jet Plane}",
    howpublished   = "{ATL-PHYS-PUB-2023-017}",
    url            = "https://cds.cern.ch/record/2864131",
    year           = "2023",
}

@Booklet{ATL-PHYS-PUB-2023-020,
    author         = "{ATLAS Collaboration}",
    title          = "{Constituent-Based \(W\)-boson Tagging with the ATLAS Detector}",
    howpublished   = "{ATL-PHYS-PUB-2023-020}",
    url            = "https://cds.cern.ch/record/2860189",
    year           = "2023",
}

@Booklet{ATL-PHYS-PUB-2023-032,
    author         = "{ATLAS Collaboration}",
    title          = "{Constituent-Based Quark Gluon Tagging using Transformers with the ATLAS detector}",
    howpublished   = "{ATL-PHYS-PUB-2023-032}",
    url            = "https://cds.cern.ch/record/2878932",
    year           = "2023",
}

@article{Salam:2010nqg,
    author = "Salam, Gavin P.",
    title = "{Towards Jetography}",
    eprint = "0906.1833",
    archivePrefix = "arXiv",
    primaryClass = "hep-ph",
    doi = "10.1140/epjc/s10052-010-1314-6",
    journal = "Eur. Phys. J. C",
    volume = "67",
    pages = "637--686",
    year = "2010"
}

@article{Komiske:2018cqr,
    author = "Komiske, Patrick T. and Metodiev, Eric M. and Thaler, Jesse",
    title = "{Energy Flow Networks: Deep Sets for Particle Jets}",
    eprint = "1810.05165",
    archivePrefix = "arXiv",
    primaryClass = "hep-ph",
    reportNumber = "MIT-CTP 5064",
    doi = "10.1007/JHEP01(2019)121",
    journal = "JHEP",
    volume = "01",
    pages = "121",
    year = "2019"
}

@misc{rikab_top_qcd,
  author       = {Gambhir, Rikab and LeBlanc, Matt and Zhou, Yuanchen},
  title        = {Pythia8 and Herwig7 Boosted Top \& QCD Jet datasets},
  month        = 8,
  year         = 2025,
  howpublished = {Zenodo},
  doi          = {10.5281/zenodo.16986897},
  url          = {https://doi.org/10.5281/zenodo.16986897}
}

@misc{komiske_2019_3164691,
  author       = {Komiske, Patrick T. and Metodiev, Eric M. and
                  Thaler, Jesse},
  title        = {Pythia8 Quark and Gluon Jets for Energy Flow, v1},
  month        = may,
  year         = 2019,
  howpublished = {Zenodo},
  doi          = {10.5281/zenodo.3164691},
  url          = {https://doi.org/10.5281/zenodo.3164691},
  note          = {https://doi.org/10.5281/zenodo.3164691}
}

@misc{pathak_2019_3066475,
  author       = {Pathak, Aditya and
                  Komiske, Patrick T. and Metodiev, Eric M. and
                  Schwartz, Matthew},
  title        = {Herwig7.1 Quark and Gluon Jets, v1},
  month        = may,
  year         = 2019,
  howpublished    = {Zenodo},
  doi          = {10.5281/zenodo.3066475},
  url          = {https://doi.org/10.5281/zenodo.3066475},
  note         = {https://doi.org/10.5281/zenodo.3066475}
}

@Article{Sjostrand:2007gs,
	author    = "Sj{\"o}strand, T. and Mrenna, S. and Skands, P.",
	title     = "{A brief introduction to PYTHIA 8.1}",
	journal   = "Comput. Phys. Commun.",
	volume    = "178",
	year      = "2008",
	pages     = "852-867",
	eprint    = "0710.3820",
	archivePrefix = "arXiv",
	primaryClass  =  "hep-ph",
	doi       = "10.1016/j.cpc.2008.01.036",
	SLACcitation  = "%%CITATION = 0710.3820;%%"
}

@article{Bellm:2017bvx,
      author         = "Bellm, Johannes and others",
      title          = "{Herwig 7.1 Release Note}",
      year           = "2017",
      eprint         = "1705.06919",
      archivePrefix  = "arXiv",
      primaryClass   = "hep-ph",
      reportNumber   = "CERN-PH-TH-2017-109, CERN-TH-2017-109, MAN-HEP-2017-08,
                        UWTHPH-2017-10, IFJPAN-IV-2017-7, NIKHEF-2017-026,
                        HERWIG-2017-02, KA-TP-19-2017, MCNET-17-08, IPPP-17-40",
      SLACcitation   = "%%CITATION = ARXIV:1705.06919;%%"
}

@book{Marzani:2019hun,
    author = "Marzani, Simone and Soyez, Gregory and Spannowsky, Michael",
    title = "{Looking inside jets: an introduction to jet substructure and boosted-object phenomenology}",
    eprint = "1901.10342",
    archivePrefix = "arXiv",
    primaryClass = "hep-ph",
    doi = "10.1007/978-3-030-15709-8",
    publisher = "Springer",
    volume = "958",
    year = "2019"
}

@article{10.2307/2232324,
    author = {Artis, M. J.},
    title = "{Monetary Theory and Practice. The UK Experience}",
    journal = {The Economic Journal},
    volume = {94},
    number = {376},
    pages = {984-985},
    year = {1984},
    month = {12},
    issn = {0013-0133},
    doi = {10.2307/2232324},
    url = {https://doi.org/10.2307/2232324},
    eprint = {https://academic.oup.com/ej/article-pdf/94/376/984/27143425/ej0984.pdf},
}

@proceedings{Proceedings:2018jsb,
    author = "Andersen, J. R. and others",
    title = "{Les Houches 2017: Physics at TeV Colliders Standard Model Working Group Report}",
    eprint = "1803.07977",
    archivePrefix = "arXiv",
    primaryClass = "hep-ph",
    reportNumber = "FERMILAB-CONF-18-122-CD-T, UWTHPH-2018-5",
    month = "3",
    year = "2018"
}

@article{Soyez:2018opl,
    author = "Soyez, Gr\'egory",
    title = "{Pileup mitigation at the LHC: A theorist\textquoteright{}s view}",
    eprint = "1801.09721",
    archivePrefix = "arXiv",
    primaryClass = "hep-ph",
    doi = "10.1016/j.physrep.2019.01.007",
    journal = "Phys. Rept.",
    volume = "803",
    pages = "1--158",
    year = "2019"
}

@InProceedings{pmlr-v162-qu22b,
  title = 	 {Particle Transformer for Jet Tagging},
  author =       {Qu, Huilin and Li, Congqiao and Qian, Sitian},
  booktitle = 	 {Proceedings of the 39th International Conference on Machine Learning},
  pages = 	 {18281--18292},
  year = 	 {2022},
  editor = 	 {Chaudhuri, Kamalika and Jegelka, Stefanie and Song, Le and Szepesvari, Csaba and Niu, Gang and Sabato, Sivan},
  volume = 	 {162},
  series = 	 {Proceedings of Machine Learning Research},
  month = 	 {17--23 Jul},
  publisher =    {PMLR},
  url = 	 {https://proceedings.mlr.press/v162/qu22b.html}
}

@article{Skands:2014pea,
    author = "Skands, Peter and Carrazza, Stefano and Rojo, Juan",
    title = "{Tuning PYTHIA 8.1: the Monash 2013 Tune}",
    eprint = "1404.5630",
    archivePrefix = "arXiv",
    primaryClass = "hep-ph",
    reportNumber = "CERN-PH-TH-2014-069, MCNET-14-08, OUTP-14-05P",
    doi = "10.1140/epjc/s10052-014-3024-y",
    journal = "Eur. Phys. J. C",
    volume = "74",
    number = "8",
    pages = "3024",
    year = "2014"
}

@article{Larkoski:2014pca,
    author = "Larkoski, Andrew J. and Thaler, Jesse and Waalewijn, Wouter J.",
    title = "{Gaining (Mutual) Information about Quark/Gluon Discrimination}",
    eprint = "1408.3122",
    archivePrefix = "arXiv",
    primaryClass = "hep-ph",
    reportNumber = "MIT--CTP-4572, NIKHEF-2014-026",
    doi = "10.1007/JHEP11(2014)129",
    journal = "JHEP",
    volume = "11",
    pages = "129",
    year = "2014"
}

@article{Benevedes:2025nzr,
    author = "Benevedes, Sean and Thaler, Jesse",
    title = "{Frequentist uncertainties on neural density ratios with wi{\,}fi ensembles}",
    eprint = "2506.00113",
    archivePrefix = "arXiv",
    primaryClass = "hep-ph",
    reportNumber = "MIT-CTP/5874",
    doi = "10.1103/w28w-x5wh",
    journal = "Phys. Rev. D",
    volume = "112",
    number = "5",
    pages = "056024",
    year = "2025"
}

@article{neyman-pearson,
  author = {Neyman, J. and Pearson, E. S.},
  copyright = {Copyright © 1933 The Royal Society},
  issn = {02643952},
  journal = {Philosophical Transactions of the Royal Society of London. Series A, Containing Papers of a Mathematical or Physical Character},
  jstor_articletype = {primary_article},
  jstor_formatteddate = {1933},
  pages = {289--337},
  publisher = {The Royal Society},
  title = {On the Problem of the Most Efficient Tests of Statistical Hypotheses},
  url = {http://www.jstor.org/stable/91247},
  volume = 231,
  year = 1933
}

@article{fastjet,
    author = "Cacciari, Matteo and Salam, Gavin P. and Soyez, Gregory",
    title = "{FastJet User Manual}",
    eprint = "1111.6097",
    archivePrefix = "arXiv",
    primaryClass = "hep-ph",
    reportNumber = "CERN-PH-TH-2011-297",
    doi = "10.1140/epjc/s10052-012-1896-2",
    journal = "Eur. Phys. J. C",
    volume = "72",
    pages = "1896",
    year = "2012"
}

@article{JETM-2023-06,
    author = "{ATLAS Collaboration}",
    title = "{Accuracy versus precision in boosted top tagging with the ATLAS detector}",
    eprint = "2407.20127",
    archivePrefix = "arXiv",
    primaryClass = "hep-ex",
    reportNumber = "CERN-EP-2024-159",
    doi = "10.1088/1748-0221/19/08/P08018",
    journal = "JINST",
    volume = "19",
    number = "08",
    pages = "P08018",
    year = "2024"
}

@misc{hinton2015distillingknowledgeneuralnetwork,
      title={Distilling the Knowledge in a Neural Network}, 
      author={Geoffrey Hinton and Oriol Vinyals and Jeff Dean},
      year={2015},
      eprint={1503.02531},
      archivePrefix={arXiv},
      primaryClass={stat.ML},
      url={https://arxiv.org/abs/1503.02531}, 
}

@article{Butter:2022xyj,
    author = "Butter, Anja and Dillon, Barry M. and Plehn, Tilman and Vogel, Lorenz",
    title = "{Performance versus resilience in modern quark-gluon tagging}",
    eprint = "2212.10493",
    archivePrefix = "arXiv",
    primaryClass = "hep-ph",
    doi = "10.21468/SciPostPhysCore.6.4.085",
    journal = "SciPost Phys. Core",
    volume = "6",
    pages = "085",
    year = "2023"
}

@misc{kingma2017adammethodstochasticoptimization,
      title={Adam: A Method for Stochastic Optimization}, 
      author={Diederik P. Kingma and Jimmy Ba},
      year={2017},
      eprint={1412.6980},
      archivePrefix={arXiv},
      primaryClass={cs.LG},
      url={https://arxiv.org/abs/1412.6980}, 
}

@inbook{10.1007/978-3-540-88908-3_14,
author = {Zitzler, Eckart and Knowles, Joshua and Thiele, Lothar},
title = {Quality Assessment of Pareto Set Approximations},
year = {2008},
isbn = {9783540889076},
publisher = {Springer-Verlag},
address = {Berlin, Heidelberg},
url = {https://doi.org/10.1007/978-3-540-88908-3\_14},
booktitle = {Multiobjective Optimization: Interactive and Evolutionary Approaches},
pages = {373–404},
numpages = {32}
}

@article{CMS:2023lpp,
    author = "{CMS Collaboration}",
    title = "{Measurement of the primary Lund jet plane density in proton-proton collisions at $ \sqrt{\textrm{s}} $ = 13 TeV}",
    eprint = "2312.16343",
    archivePrefix = "arXiv",
    primaryClass = "hep-ex",
    reportNumber = "CMS-SMP-22-007, CERN-EP-2023-282",
    doi = "10.1007/JHEP05(2024)116",
    journal = "JHEP",
    volume = "05",
    pages = "116",
    year = "2024"
}

@article{ATLAS:2024wrd,
    author = "{ATLAS Collaboration}",
    title = "{Measurements of Lund subjet multiplicities in 13 TeV proton-proton collisions with the ATLAS detector}",
    eprint = "2402.13052",
    archivePrefix = "arXiv",
    primaryClass = "hep-ex",
    reportNumber = "CERN-EP-2024-029",
    doi = "10.1016/j.physletb.2024.139090",
    journal = "Phys. Lett. B",
    volume = "859",
    pages = "139090",
    year = "2024"
}

@article{ATLAS:2025qtv,
    author = "{ATLAS Collaboration}",
    title = "{Measurement of jet track functions in pp collisions at s=13 TeV with the ATLAS detector}",
    eprint = "2502.02062",
    archivePrefix = "arXiv",
    primaryClass = "hep-ex",
    reportNumber = "CERN-EP-2024-333",
    doi = "10.1016/j.physletb.2025.139680",
    journal = "Phys. Lett. B",
    volume = "868",
    pages = "139680",
    year = "2025"
}

@article{51791361-8fe2-38d5-959f-ae8d048b490d,
 ISSN = {00359246},
 URL = {http://www.jstor.org/stable/2346178},
 author = {Robert Tibshirani},
 journal = {Journal of the Royal Statistical Society. Series B (Methodological)},
 number = {1},
 pages = {267--288},
 publisher = {[Royal Statistical Society, Oxford University Press]},
 title = {Regression Shrinkage and Selection via the Lasso},
 urldate = {2025-11-19},
 volume = {58},
 year = {1996}
}

@article{a92f3c16-7c6e-31d3-b403-82d2b0a469e4,
 ISSN = {00401706},
 URL = {http://www.jstor.org/stable/1267351},
 author = {Arthur E. Hoerl and Robert W. Kennard},
 journal = {Technometrics},
 number = {1},
 pages = {55--67},
 publisher = {[Taylor & Francis, Ltd., American Statistical Association, American Society for Quality]},
 title = {Ridge Regression: Biased Estimation for Nonorthogonal Problems},
 urldate = {2025-11-19},
 volume = {12},
 year = {1970}
}

@book{tikhonov1977solutions,
  address = {Washington, D.C.: John Wiley \& Sons, New York},
  author = {Tikhonov, Andrey N. and Arsenin, Vasiliy Y.},
  description = {MR: Publications results for "MR Number=(455365)"},
  mrclass = {65J05 (65R05 65NXX)},
  mrnumber = {0455365 (56 \#13604)},
  mrreviewer = {M. Z. Nashed},
  note = {Translated from the Russian, Preface by translation editor Fritz John, Scripta Series in Mathematics},
  pages = {xiii+258},
  publisher = {V. H. Winston \& Sons},
  title = {Solutions of ill-posed problems},
  year = 1977
}

@inproceedings{Liu:2023dio,
    author = "Liu, Ryan and Gandrakota, Abhijith and Ngadiuba, Jennifer and Spiropulu, Maria and Vlimant, Jean-Roch",
    title = "{Efficient and Robust Jet Tagging at the LHC with Knowledge Distillation}",
    booktitle = "{37th Conference on Neural Information Processing Systems}",
    eprint = "2311.14160",
    archivePrefix = "arXiv",
    primaryClass = "hep-ex",
    reportNumber = "FERMILAB-PUB-23-748-CMS",
    month = "11",
    year = "2023"
}

}

%%%%%%%%%%%%%%%%%%%%%%%%%%%%%%%%%%%%%%%%%%%%%%%%%%%%%%%%%%%%%%%%%%%%%%%%%%%%%

\clearpage
% \appendix

% \section{Appendix: Knowledge Distillation}

% Lorem 

% \section{Appendix: Quark/Gluon Fraction Case Study}

% Lorem

% \section{Appendix: Public top tagging samples}

% Lorem

\end{document}